\documentclass[a4paper,11pt]{article}
\pdfoutput=1 % to allow compilation in JCAP
\usepackage{jcappub}
\usepackage{amsmath,amssymb}
\usepackage{bm}
\usepackage{xcolor}
\usepackage{graphicx}
\usepackage{enumitem}
\usepackage{geometry}
\usepackage{xspace}
\usepackage{pdflscape}
\usepackage{placeins}
\usepackage{epstopdf}
\usepackage{comment}
\makeatletter
\gdef\@fpheader{}
\g@addto@macro\bfseries{\boldmath}
\makeatother

\newcommand{\OmegaGW}{\Omega_{\mathrm{GW}}}
\newcommand{\rhoGW}{\rho_{\mathrm{GW}}}

\let\oldsqrt\sqrt
\def\sqrt{\mathpalette\DHLhksqrt}
\def\DHLhksqrt#1#2{%
\setbox0=\hbox{$#1\oldsqrt{#2\,}$}\dimen0=\ht0
\advance\dimen0-0.2\ht0
\setbox2=\hbox{\vrule height\ht0 depth -\dimen0}%
{\box0\lower0.4pt\box2}}

\newcommand{\sss}[1]{{\scriptscriptstyle{#1}}}
\newcommand{\boldmathsymbol}[1]{{\ensuremath{\boldsymbol{#1}}}}

\newcommand{\uPl}{\mathrm{Pl}}

\newcommand{\usssPl}{\sss{\uPl}}

\newcommand{\calH}{\mathcal{H}}

\newcommand{\Mp}{M_\usssPl}

\newcommand{\beq}{\begin{equation}}
\newcommand{\eeq}{\end{equation}}
\newcommand{\bea}{\begin{equation}\begin{aligned}}
\newcommand{\eea}{\end{aligned}\end{equation}}

\newlength{\wsingfig}
\newlength{\wdblefig}
\newlength{\wquadfig}
\newlength{\wtriplefig}
\newcommand{\Eq}[1]{Eq.~(\ref{#1})}

\newcommand{\Fig}[1]{Fig.~{\ref{#1}}}

\newcommand{\Sec}[1]{Sec.~\ref{#1}}

\newcommand{\Hc}[1]{\mathcal{H}}
\renewcommand{\Hc}{\mathcal{H}}

\def\doi{http://doi.org}

\date{today}

\title{Gravitational-wave signatures of gravito-electromagnetic couplings}

\author[a,b,c]{Theodoros~Papanikolaou}
\author[d,c]{Charalampos~Tzerefos}
\author[a,b]{Salvatore~Capozziello}
\author[e,f]{Gaetano~Lambiase}

\affiliation[a]{Scuola Superiore Meridionale, Largo San Marcellino 10, 80138 Napoli, Italy}

\affiliation[b]{Istituto Nazionale di Fisica Nucleare (INFN), Sezione di Napoli, Via Cinthia 21, 80126 Napoli, Italy}

\affiliation[c]{National Observatory of Athens, Lofos Nymfon, 11852 Athens, 
Greece}

\affiliation[d]{Department of Physics, National \& Kapodistrian University of Athens, Zografou Campus GR 157 73, Athens, Greece}

\affiliation[e]{Dipartimento di Fisica, Universita` di Salerno, Via Giovanni Paolo II, 132 I-84084 Fisciano (SA), Italy}

\affiliation[f]{INFN, Sezione di Napoli, Gruppo collegato di Salerno, Italy}

\emailAdd{t.papanikolaou@ssmeridionale.it}
\emailAdd{chtzeref@phys.uoa.gr}
\emailAdd{capozziello@na.infn.it}
\emailAdd{lambiase@sa.infn.it}

\abstract{Gravitational waves (GWs) can undoubtedly serve as a messenger from the early Universe acting as well as a novel probe of the underlying gravity theory. In this work, motivated by one-loop vacuum-polarization effects on curved spacetime, we investigate a gravitational theory with non-minimal curvature-electromagnetic coupling terms of the form $\xi R F_{\mu\nu}F^{\mu\nu}$, where $R$ is the scalar curvature and $F_{\mu\nu}$ the Faraday tensor, which can be responsible for the generation of primordial electromagnetic fields. We study then the GW signatures of such coupling terms deriving in particular for the first time to the best of our knowledge the modified tensor modes equation of motion. Remarkably, we find a universal infrared (IR) frequency scaling $f^5$ of the electromagnetically induced GW (EMIGW) signal, which, depending on the energy scale of inflation, the duration of inflation and reheating as well as the dynamical behaviour of the gauge coupling function $\xi$, can be well within the detection sensitivity bands of GW experiments such as SKA, LISA, ET and BBO, being thus potentially detectable in the future by GW observatories.
}

\begin{document}

\maketitle

%%%%%%%%%%%%%%%%%%%%%%%% Section 1:  Introduction %%%%%%%%%%%%%%%%%%%%%%%%%%%%%%%%%%%%%%%%%%%%%%%%%%%%%%%%%%%%%%%%
\section{Introduction} 

The detection of a gravitational wave (GW) signal emitted from a binary black-hole merger ten years ago by the LIGO/VIRGO collaboration~\cite{LIGOScientific:2016aoc} inaugurated the era of GW astronomy. Remarkably, through the portal of gravitational waves one can probe sources and cosmic epochs which are not accessible through the classical electromagnetic radiation based astronomy. In particular, the GW portal can serve as a messenger from the early Universe, where one is met with numerous GW production mechanisms, namely inflation~\cite{Rubakov:1982df,Starobinsky:1983zz}, phase transitions~\cite{1986MNRAS.218..629H,Kosowsky:1992vn,Kosowsky:1992rz,Kamionkowski:1993fg,Caprini:2007xq,Huber:2008hg}, topological defects~\cite{Durrer:1998rw,Dunsky:2021tih,Hiramatsu:2010yz}, primordial black holes~\cite{LISACosmologyWorkingGroup:2023njw} and primordial turbulence~\cite{Kahniashvili:2005qi,Caprini:2006jb,Gogoberidze:2007an,Kahniashvili:2008er,Caprini:2009yp}.

Furthermore, magnetic fields (MFs) are pervasively observed in the Universe over a wide range of scales affecting unavoidably its dynamical evolution and being accompanied with a very rich phenomenology. Interestingly enough, MFs can significantly influence the process of particle acceleration through the intergalactic medium~\cite{Bagchi:2002vf} and the propagation of cosmic rays~\cite{Strong:2007nh} while at the same time entwined with GWs~\cite{Addazi:2024osi,Addazi:2024kbq}, MFs can give us a multi-messenger access to the early Universe. However, their generation mechanism is still unknown and constitutes an active field of research. In particular, numerous cosmological MF generation mechanisms have been proposed, ones associated to phase transitions~\cite{1988PhRvD..37.2743T,1989ApJ...344L..49Q}, primordial scalar~\cite{Ichiki:2006cd,Naoz:2013wla,Flitter:2023xql} and vector perturbations~\cite{Banerjee:2003xk,Durrer:2006pc} as well as astrophysical ones seeding battery-induced MFs~\cite{1950ZNatA...5...65B,Widrow:2002ud,Hanayama:2005hd,Safarzadeh:2017mdy,Araya:2020tds,Papanikolaou:2023nkx}.

One of the most important MF generation mechanisms is the inflationary one, via which primordial magnetic fields coherent on large length scales naturally arise ~\cite{Turner:1987bw,Ratra:1991av,Bamba:2006ga,Martin:2007ue}. The most studied phenomenological model to achieve such MF generation is the one firstly introduced by Ratra~\cite{Ratra:1991bn}, in which a time-dependent function, usually a function of the inflaton field, is coupled to the usual electromagnetic (EM) action. Despite the fact that this MF generation mechanism produces coherent magnetic fields with strength compatible with MF observational constraints, it suffers from the back-reaction and strong coupling problems~\cite{Demozzi:2009fu}, related to unnaturally strong couplings between charged particles and the EM field as well to extremely high EM energy densities larger than that of the background. However, these issues can be circumvented with a suitable choice of the time-dependent function coupled to the EM action~\cite{Sharma:2017eps,BazrafshanMoghaddam:2017zgx,Sharma:2018kgs,Sobol:2018djj,Kobayashi:2019uqs}.

Motivated thus by inflationary magnetogenesis and the access gained to the early Universe through the GW portal, we study in this work the generation of GWs produced during the post-inflationary era, which are sourced by gravitational-electromagnetic couplings responsible for the generation of primordial magnetic fields. In contrast to previous works on the topic~\cite{Sharma:2019jtb,Brandenburg:2021pdv,Brandenburg:2021bfx,RoperPol:2022iel} concerning couplings of the inflaton field with the kinetic electromagnetic Lagrangian, we will focus here on GWs induced by gravitational-electromagnetic couplings which can naturally arise within the quantum field theory (QFT) description of massless U(1) gauge fields, responsible for electromagnetism, on curved spacetime. In particular, at the level of the Lagrangian we will consider coupling terms of the form $\mathcal{R}^nF^{\mu\nu}F_{\mu\nu}/\Mp^{2n}$~\cite{DeWitt:1964mxt}, with $\Mp$ being the reduced Planck mass, $\mathcal{R}$ any component of the Riemann tensor and $n$ a free parameter. These terms can actually naturally arise due to one-loop vacuum-polarization effects in curved spacetime~\cite{Drummond:1979pp,Turner:1987bw} as well as  within the context of string-theory setups~\cite{Cai:1986sa,Hegarty:1992sj}.

%Apart from $\mathcal{R}F^{\mu\nu}F_{\mu\nu}/\Mp^2$ terms, one can consider as well CP-violating dimension-six Chern-Simons terms of the form $RF_{\mu\nu}\tilde{F}^{\mu\nu}$ being responsible for the generation of large-scale primordial magnetic fields without suffering from strong coupling and bakcreaction issues~\cite{Subramanian:2015lua,Savchenko:2018pdr,Durrer:2022emo,Durrer:2023rhc} as well as for the baryon-assymetry in the Universe~\cite{Cado:2023zbm}. However, as we will see in the following these last terms terms will lead to negligible gravitational-wave signals.

%Furthermore, we will study as well the effect of the underlying gravity theory on the aforementioned magnetically induced GW signal. In particular, we will consider theories where the gravity Lagrangian is promoted to a  function of the scalar curvature $R$, namely $f(R)$ gravity theories, where one is unavoidably met with an extra GW polarisation mode compared to the general relativity (GR) case, promoting thus the portal of GWs induced by gravity - electromagnetic coupling as a new probe of the nature of gravity.

%Interestingly enough,  we find that for some characterstic viable classes of f(R) gravity, the GW propagator of the scalaron mode is the same with the of GR giving rise at the end to an overall enhancement of the GW amplitude compare to the GR case.

This paper is structured as follows: In \Sec{sec:curvature-electromagnetism coupling} we introduce a gravitational theory with a non-minimal curvature-electromagnetic coupling responsible for the generation of primordial electromagnetic fields. Then, in \Sec{sec:EM_sector} after quantising our vector gauge field and making a choice on the parametrisation of the gauge coupling function $\xi$, we extract the electric and magnetic field power spectra. We derive as well the necessary conditions in order to avoid strong coupling and back-reaction issues present at theories with non-minimal gravitational-electromagnetic couplings. Subsequently, in \Sec{sec:EMIGWs}, we derive the modified equation of motion for the tensor modes extracting as well the electromagnetically induced GW signal today and comparing it with the frequency detection sensitivity curves of GW experiments, namely that of SKA, LISA, ET and BBO. Finally, \Sec{sec:Conclusions} is devoted to conclusions.

\section{Introducing a curvature-electromagnetic coupling}\label{sec:curvature-electromagnetism coupling}
We consider here a gravitational theory where gravity is coupled with electromagnetism. Notably, within the context of Quantum Electrodynamics (QED) on curved spacetime, one can perform a Schwinger-DeWitt expansion (SDWE) expansion in powers of $\mathcal{R}/\Mp^2$~\cite{DeWitt:1964mxt}, where $\Mp$ is the reduced Planck mass and $\mathcal{R}$ is any component of the Riemann tensor, and write an effective electromagnetic action containing couplings of the form $\xi(t)\frac{\mathcal{R}^n}{M^{2n}_\mathrm{Pl}}F_{\mu\nu}F^{\mu\nu}$~\cite{Mazzitelli:1995mp}, with $\xi$ being a coupling function which in general can be a function of time~\cite{Bamba:2008ja}. 
%In addition, recently it has been noticed that CP-violating dimension-six Chern-Simons terms of the form $RF_{\mu\nu}\tilde{F}^{\mu\nu}$ can naturally lead to baryogenesis, being responsible as well for the generation of large-scale primordial magnetic without suffering from strong coupling and backreaction issues~\cite{Subramanian:2015lua,Savchenko:2018pdr,Durrer:2022emo,Durrer:2023rhc}
For the purposes of our work, for simplicity we will consider only the $n=1$ term coupling only the scalar curvature with the electromagnetic sector, i.e. the term $R F_{\mu\nu}F^{\mu\nu}$. Therefore, the total action of our physical system at hand will be recast as
\beq\label{eq:S_tot}
\begin{split}
S = & \frac{1}{16\pi G} \int d^4x \sqrt{-g}R - \frac{1}{4}\int d^4x \sqrt{-g} F_{\mu\nu}F^{\mu\nu} \\ & - \frac{1}{\Mp^2}\int d^4x \sqrt{-g} \xi(t) RF_{\mu\nu}F^{\mu\nu} 
+ \int d^4x \sqrt{-g} \mathcal{L}_\mathrm{m} ,
\end{split}
\eeq
where $F_{\mu\nu}$ is the electromagnetic (Faraday) tensor written in terms of the vector potential $A_\mu$ as  $F_{\mu\nu}\equiv \partial_\mu A_\nu - \partial_\nu A_\mu$. %and $\tilde{F}_{\mu\nu}$ is its Hodge dual, which in our case is equal to $\tilde{F}_{\mu\nu}= \frac{1}{2} \epsilon_{\mu \nu \alpha \beta} F^{\alpha \beta} $, with $\epsilon_{\mu \nu \alpha \beta}$ being the fully anti-symmetric Levi-Civita symbol.
The second term describes standard electromagnetism on curved spacetime, while the third one corresponds to a non-minimal coupling term of the scalar curvature with the kinetic electromagnetic Lagrangian. $\mathcal{L}_\mathrm{m}$ stands for the Lagrangian density of the rest of the matter sector which does not depend on the electromagnetic (EM) fields.

The next step is the  derivation of the gravitational field equations, which is accomplished via varying the action \eqref{eq:S_tot} with respect to $\delta g^{\mu \nu}$, leading to     
\bea 
\label{eq: GravField equation}
& R_{\mu \nu} - \frac{1}{2} g_{\mu \nu} R = \kappa^2 T ^{(em)}_{\mu \nu}, \, \, \text{with} \, \, \\
& T ^{(em)}_{\mu \nu} \equiv (1 + \xi R ) \left( g^{\alpha \beta} F_{\mu \beta} F_{\nu \alpha} - \frac{1}{4} g_{\mu \nu} F_{\alpha \beta} F^{\alpha \beta} \right) %+ \xi_2 R \left( g^{\alpha \beta} F_{\mu \beta} \tilde{F}_{\nu \alpha} - \frac{1}{4} g_{\mu \nu} F_{\alpha \beta} \tilde{F}^{\alpha \beta} \right) 
\\ & + \frac{\xi }{2}\left[ R_{\mu \nu} F_{\alpha \beta} F^{\alpha \beta} + g_{\mu \nu} \Box \left( F_{\alpha \beta} F^{\alpha \beta}\right) - \nabla_{\mu}\nabla_{\nu} \left( F_{\alpha \beta} F^{\alpha \beta}\right)  \right], % \\ & + \frac{\xi_2 }{2}\left[ R_{\mu \nu} F_{\alpha \beta} \tilde{F}^{\alpha \beta} + g_{\mu \nu} \Box \left( F_{\alpha \beta}\tilde{F}^{\alpha \beta}\right) - \nabla_{\mu}\nabla_{\nu} \left( F_{\alpha \beta} \tilde{F}^{\alpha \beta}\right)  \right]
\eea
where $\kappa ^2 \equiv 1/\Mp^2$  \footnote{Throughout the paper the metric signature used is $(-,+,+,+)$.
}.

We observe here that the left hand side is the familiar from general relativity Einstein tensor, while on the right hand side we have the energy momentum tensor of standard electromagnetism on curved spacetime ($T_{\mu \nu}= g^{\alpha \beta} F_{\mu \beta} F_{\nu \alpha} - \frac{1}{4} g_{\mu \nu} F_{\alpha \beta} F^{\alpha \beta} $ ), albeit it has one extra coupling function and some extra terms arising from the direct gravito-electromagnetic coupling. 

\section{The electromagnetic dynamics}\label{sec:EM_sector}

Let us now derive the dynamics of the EM sector. Specifically, by varying \Eq{eq:S_tot} with respect to the EM tensor $F_{\mu\nu}$, one gets the Euler-Lagrange equation for $A_i$, which can be recast as $\partial_\mu\left[\sqrt{-g}\left(1+\xi R\right)g^{\mu\lambda}g^{\nu\rho}F_{\lambda\rho}\right]=0$. For a cosmic epoch with $w=p/\rho$, one can show that the Ricci scalar satisfies $R=6(2H^2+\dot{H}) = (1-3w)\rho/\Mp^2$, which for $w<1/3$ is a positive definite quantity. Thus, in order to simplify the calculation and being inspired by the analysis performed for the Ratra coupling of the form $f^2(\phi)F^{\mu\nu}F_{\mu\nu}$~\cite{Ratra:1991bn}, we introduce the function $f(t)$ defined as 
\beq\label{eq:f_definition}
f^2(t) \equiv 1+\xi(t)R(t),
\eeq
with $\xi>0$ in order to avoid strong coupling issues~\cite{Durrer:2022emo}~\footnote{Since the effective electron charge is defined as $e_\mathrm{eff}\equiv e/f^2$ and for the metric signature $(-,+,+,+)$ used throughout the paper, $R$ is positive, $\xi$ should be positive as well so as to avoid the presence of singular points in $f^2$, thus unnaturally large values of $e_\mathrm{eff}$.}. After a straightforward calculation and working in the Coulomb gauge, i.e. $A_0 =0$ and $\partial_iA^i = 0$, one can show that the equation of motion for $A_i$ will read as
\beq\label{eq:MS_A_i_cosmic_time}
\ddot{A}_i(\boldmathsymbol{x},t) + \left(H+2\frac{\dot{f}}{f}
\right)\dot{A}_i(\boldmathsymbol{x},t) - \nabla^2 A_i(\boldmathsymbol{x},t)=0,
\eeq
where a dot denotes differentiation with respect to the cosmic time. In terms of the conformal time one can recast the above equation as
\beq\label{eq:MS_A_i_conformal_time}
A^{\prime\prime}_i(\boldmathsymbol{x},\eta) + 2\frac{f^\prime}{f}A^\prime_i(\boldmathsymbol{x},\eta) - a^2\nabla^2 A_i(\boldmathsymbol{x},\eta)=0,
\eeq
where $\nabla^2\equiv g^{ij}\partial_i\partial_j$. 

\subsection{The quantisation of the vector gauge field}
Quantising now the $U(1)$ gauge field $A_i(\boldmathsymbol{x},\eta)$ we decompose it in terms of creation 
and annihilation operators, namely $\hat{b}(\boldmathsymbol{k})^\dagger$ and $\hat{b}(\boldmathsymbol{k})$ respectively. In particular, one can write $A_i(\eta,\boldmathsymbol{x})$ as 
\beq
\begin{split}
A_i(\boldmathsymbol{x},\eta) &  =  \int\frac{\mathrm{d}^3k}{(2\pi)^{3/2}}\sum_{\sigma = 1,2}\Bigl[\hat{b}(\boldmathsymbol{k},\sigma)\epsilon_i(\boldmathsymbol{k},\sigma)A(k,\eta)e^{i\boldmathsymbol{k}\cdot\boldmathsymbol{x}} \\ & +\hat{b}^\dagger(\boldmathsymbol{k},\sigma)\epsilon^{*}_i(\boldmathsymbol{k},\sigma)A^{*}(k,\eta)e^{-i\boldmathsymbol{k}\cdot\boldmathsymbol{x}}\Bigr],
\end{split}
\eeq
where $\sigma = (1,2)$ stands for the two polarisation states of the electromagnetic field and  $\epsilon_i(\boldmathsymbol{k},\sigma)$ are the transverse polarization vectors satisfying the completeness relation $\sum_\sigma \epsilon_i(\boldmathsymbol{k},\sigma) \epsilon^{*}_j(\boldmathsymbol{k},\sigma) = P_{ij}(\boldmathsymbol{k})$, with $P_{ij}(\boldmathsymbol{k}) = \delta_{ij} - k_ik_j/k^2$. The annihilation and creation operators $\hat{b}$ and $\hat{b}^\dagger$ should satisfy the usual commutation relations $[\hat{b}(\boldmathsymbol{k},\sigma),\hat{b}^\dagger(\boldmathsymbol{k}^\prime,\sigma^\prime)] = (2\pi)^3\delta_{\sigma\sigma^\prime}\delta(\boldmathsymbol{k}-\boldmathsymbol{k}^\prime)$ implying the normalisation condition 
\beq\label{eq:Wronskian_A_k}
A(k,\eta)A^{\prime,*}(k,\eta) - A^{*}(k,\eta)A^\prime(k,\eta) = \frac{i}{f^2(\eta)a^3(\eta)}
\eeq
for the Fourier modes of the gauge field. One then can show that the gauge field $A_i$ and its conjugate momentum $\pi_j \equiv \delta S/ \delta \dot{A}^j$ satisfy the usual commutation relation
\beq\label{eq:A_i_normalisation}
[A_i(\eta,\boldmathsymbol{x}),\pi_j(\eta,\boldmathsymbol{y})] = i\int\frac{d^3\mathrm{k}}{(2\pi)^3}e^{i\boldmathsymbol{k}\cdot(\boldmathsymbol{x}-\boldmathsymbol{y})}P_{ij}(\boldmathsymbol{k}).
\eeq
At the end, introducing the new variable $\mathcal{A}(\eta,k)\equiv a(\eta)f(\eta)A(\eta,k)$ one can write \Eq{eq:MS_A_i_conformal_time} in Fourier space in the following compact form:
\beq\label{eq:MS_A_k_Fourier_space}
\mathcal{A}^{\prime\prime}(k,\eta) + \left(k^2-\frac{f^{\prime\prime}}{f}\right)\mathcal{A}(k,\eta) = 0.
\eeq
which is the harmonic oscillator equation but with a spacetime-dependent frequency. 
%%%%%%%%%%%%%%%%%%%%%%%%%%%%%%%%%%%%%%%%%%%%%%%%%%%%%%%%%%%
\subsection{The choice of the coupling function}
In order to solve \Eq{eq:MS_A_k_Fourier_space}, we need to make a choice about the dependence of $f$, or equivalently of $\xi$, on time or on the scale factor. In order to account for the strong-coupling issue, present in models of inflationary magnetogenesis~\cite{Kobayashi:2014zza,Sharma:2017eps}, one should assume a functional form of $f$ such that the effective electron-charge defined as $e_\mathrm{eff}\equiv e/f^2$ goes to the bare electron charge $e$ at the end of the reheating phase, so as to recover classical electromagnetism during the radiation-dominated (RD) era. Consequently, we will adopt for $\xi$ a broken power-law dependence on the scale factor $a$~\cite{Bamba:2006ga,Martin:2007ue,Durrer:2022emo} recasting it as
\bea\label{eq:xi_parametrisation}
\xi =
\begin{cases}
    \left(\frac{a}{a_\mathrm{i}}\right)^{\nu_1}, \, &a_\mathrm{i}\leq a\leq a_\mathrm{inf} \\
    \left(\frac{a_\mathrm{inf}}{a_\mathrm{i}}\right)^{\nu_1}\left(\frac{a}{a_\mathrm{inf}}\right)^{\nu_2}, \, &a_\mathrm{inf}\leq a\leq a_\mathrm{reh},
\end{cases}
\eea
where $a_\mathrm{i}$ and $a_\mathrm{inf}$ stand for the scale factor at the beginning and the end of inflation and $a_\mathrm{reh}$ for the scale factor at reheating (the end of the post-inflationary era). In the following, we normalise $a_\mathrm{i}= 1$ and impose continuity of the scale factor at the end of inflation. Therefore, the scale factor will read as 

\bea\label{eq:a_parametrisation}
a =
\begin{cases}
   -\frac{1}{H_\mathrm{inf}\eta}, \, &\eta_\mathrm{i}\leq \eta \leq \eta_\mathrm{inf}, \quad \mathrm{(de \,\, Sitter\;expansion)} \\
   \frac{a^3_\mathrm{inf}H^2_\mathrm{f}}{4}\left(\frac{3}{a_\mathrm{inf}H_\mathrm{inf}} + \eta\right)^2, \, &\eta_\mathrm{inf}\leq \eta \leq \eta_\mathrm{reh},
\end{cases}
\eea
where $\eta$ stands for the conformal time.

At the end, since $R^{1/2}\approx H $ and the Hubble parameter is almost constant during inflation (quasi-de Sitter expansion),  while during reheating decays as $H\propto a^{-3/2}$ (matter-dominated era), one gets that 
\bea\label{eq:f_parametrisation}
f \simeq
\begin{cases}
    \left(\frac{a}{a_\mathrm{i}}\right)^{\alpha}\frac{H_\mathrm{inf}}{\Mp} , \, &a_\mathrm{i}\leq a\leq a_\mathrm{inf} \\
    \left(\frac{a_\mathrm{inf}}{a_\mathrm{i}}\right)^{\alpha}\frac{H_\mathrm{inf}}{\Mp}\left(\frac{a}{a_\mathrm{inf}}\right)^{-\beta}, \, &a_\mathrm{inf}\leq a\leq a_\mathrm{reh},
\end{cases}
\eea
where we introduced the new parameters $\alpha$ and $\beta$ defined as 
\beq
\alpha \equiv \nu_1/2,\quad \beta \equiv (3-\nu_2)/2.
\eeq

\subsection{The dynamics of the vector gauge field}
Imposing then the Bunch-Davies vacuum initial condition on subhorizon scales for $\mathcal{A}(k,\eta)$, i.e. $\mathcal{A}(k\gg aH,\eta)= \frac{1}{\sqrt{2k}}e^{-ik\eta}$~\footnote{Note here that the Bunch-Davies vacuum normalisation at the level of $\mathcal{A}(\eta,k)$ reproduces correctly the Wronskian given by \Eq{eq:Wronskian_A_k}.}, one can write the solution of \Eq{eq:MS_A_k_Fourier_space} during inflation in terms of Bessel functions as
\beq\label{eq:A_inf}
\mathcal{A}_\mathrm{inf}(k,\eta) = \sqrt{-k\eta}\left[d_1(k)J_{-\alpha-\frac{1}{2}}(-k\eta) + d_2(k)J_{\alpha+\frac{1}{2}}(-k\eta)\right],
\eeq
with 
\beq
d_1 =\sqrt{\frac{\pi}{4k}}\frac{e^{\frac{i\pi\alpha}{2}}}{\cos(-\pi\alpha)}, \quad d_2 = \sqrt{\frac{\pi}{4k}}\frac{e^{\frac{i\pi(-\alpha+1)}{2}}}{\cos(-\pi\alpha)}.
\eeq
In the superhorizon regime, one gets from \Eq{eq:A_inf} that 
\beq\label{eq:A_inf_superhorizon}
\begin{split}
\mathcal{A}_\mathrm{inf}(k\ll aH,\eta)  \simeq & (-k\eta)^{-\alpha} \left[\frac{2^{\frac{1}{2}+\alpha}d_1}{\Gamma\left(\frac{1}{2}-\alpha\right)}  - \frac{2^{-\frac{3}{2}+\alpha}d_1k^2\eta^2}{\Gamma\left(\frac{3}{2}-\alpha\right)} + \frac{2^{-\frac{9}{2}+\alpha}d_1k^4\eta^4}{\Gamma\left(\frac{5}{2}-\alpha\right)} + O(k^5)\right] 
\\ & + (-k\eta)^{1+\alpha}\left[\frac{2^{-\frac{1}{2}-\alpha}d_2}{\Gamma\left(\frac{3}{2}+\alpha\right)} - \frac{2^{-\frac{5}{2}-\alpha}d_2k^2\eta^2}{\Gamma\left(\frac{5}{2}+\alpha\right)} + \frac{2^{-\frac{11}{2}-\alpha}d_2k^4\eta^4}{\Gamma\left(\frac{7}{2}+\alpha\right)} + O(k^5) \right].
\end{split}
\eeq
With this solution at hand, we shall proceed with the post-inflationary era.
Focusing again on superhorizon scales $k\ll aH$, one can neglect the $k^2$ term in \Eq{eq:MS_A_k_Fourier_space} (WKB approximation) and write the solution for $\bar{A}(k,\eta)$ during reheating, defined as $\bar{A}(k,\eta)\equiv \mathcal{A}_\mathrm{reh}(k,\eta)/f_\mathrm{reh}$, as~\cite{Bamba:2006ga}
\beq
\bar{A}_\mathrm{reh}(k\ll aH,\eta) =  D_1(k) + D_2(k)\int_{\eta_\mathrm{inf}}^\eta\frac{1}{f^2_\mathrm{reh}(\bar{\eta})}\mathrm{d}\bar{\eta},
\eeq
where $D_1(k)$ is a decaying mode for $f_\mathrm{reh}$ being a decreasing function of time and $D_2$ a growing one. 

Imposing finally continuity at the end of inflation for the quantity $\bar{A}(k,\eta)\equiv \mathcal{A}(k,\eta)/f$ and its first time derivative, one can obtain the dominant terms in the superhorizon regime for the $D_1$ and $D_2$ factors reading as
\begin{align}
D_1(k) & \simeq \left(\frac{k}{H_\mathrm{inf}}\right)^{-\alpha}\Biggl[\frac{2^{\frac{1}{2}+\alpha}d_1}{\Gamma\left(\frac{1}{2}-\alpha\right)}  - \frac{2^{-\frac{3}{2}+\alpha}d_1k^2}{a^2_\mathrm{inf}H^2_\mathrm{inf}\Gamma\left(\frac{3}{2}-\alpha\right)} + \frac{2^{-\frac{9}{2}+\alpha}d_1k^4}{a^4_\mathrm{inf}H^4_\mathrm{inf}\Gamma\left(\frac{5}{2}-\alpha\right)} \\ & + \frac{2^{-\frac{1}{2}-\alpha}d_2\left(\frac{k}{a_\mathrm{inf}H_\mathrm{inf}}\right)^{1+2\alpha}}{\Gamma\left(\frac{3}{2}+\alpha\right)}\Biggr], \\
D_2(k) & \simeq \Biggl\{\frac{2^{-\frac{5}{2}-\alpha}}{\Mp^2}a^{2\alpha-3}_\mathrm{inf}\left(\frac{k}{H_\mathrm{inf}}\right)^{1-\alpha}\Bigl[-\frac{8\alpha+4}{\Gamma\left(\frac{3}{2}+\alpha\right)}a^3_\mathrm{inf}H^3_\mathrm{inf}d_2
\left(\frac{k}{a_\mathrm{inf}H_\mathrm{inf}}\right)^{2\alpha}\\ & -\frac{2^{\alpha}}{\Gamma\left(\frac{5}{2}-\alpha\right)}d_1 k^3 + \frac{2^{2+2\alpha}}{\Gamma\left(\frac{3}{2}-\alpha\right)}d_1a^2_\mathrm{inf}H^2_\mathrm{inf}k\Bigr]\Biggr\}.
\end{align}

One then gets that the rescaled gauge field $\bar{A}(k,\eta)$ during the post-inflationary era on superhorizon scales will be recast as 
\beq\label{A_reh_superhorizon}
\bar{A}_\mathrm{reh}(k\ll aH,\eta) \simeq D_1(k) + \frac{D_2(k)}{a_\mathrm{inf}H_\mathrm{inf}}\frac{\Mp^2}{H^2_\mathrm{inf}}\frac{a^{-2\alpha}_\mathrm{inf}}{(\frac{1}{2}+2\beta)}\left[\left(\frac{a_\mathrm{reh}(\eta)}{a_\mathrm{inf}}\right)^{2\beta + \frac{1}{2}} - 1\right].
\eeq

%%%%%%%%%%%%%%%%%%%%%%%%%%%%%%
\subsection{The electromagnetic field power spectrum}
Having extracting above the dynamics of the gauge field on super-horizon scales during inflation and the post-inflationary era, one can derive now the electric and the magnetic field power spectra, which will be the source of the GW signal. %~\footnote{In principle, both the electric and magnetic fields will source the production of GWs but at the end of the post-inflationary era, the electric field is damped due to the high conductivity of the thermal plasma produced at reheating~\cite{Sharma:2019jtb}.Thus, in the present paper, we will focus only on the contribution of the magnetic fields to the generation of tensor perturbations.}.
Starting with the proper electric and magnetic field, the latter can be written as~\cite{Bamba:2006ga}
\begin{align*}
E^{\mathrm{proper}}_i(t,\boldmathsymbol{x})  & = a^{-1}E_i(t,\boldmathsymbol{x}) = -a^{-1}\dot{A}_i(t,\boldmathsymbol{x}),\\
B^{\mathrm{proper}}_i(t,\boldmathsymbol{x}) &  = a^{-1}B_i(t,\boldmathsymbol{x})  = a^{-2}\epsilon_{ijk}\partial_j A_k(t,\boldmathsymbol{x}).
\end{align*}
where $\epsilon_{ijk}$ is the totally antisymmetric tensor ($\epsilon_{123}=1$). Going in Fourier space, one can straightforwardly show that electric and magnetic field amplitudes square can be recast as 
\beq
|B^{\mathrm{proper}}(k,\eta)|^2 = 2\frac{k^2}{a^4}|a A(k,\eta)|^2,\quad |E^{\mathrm{proper}}(k,\eta)|^2 = 2\frac{1}{a^4} \left\vert \frac{\partial \left(a A(k,\eta)\right)}{\partial\eta}\right\vert ^2.
\eeq
where the factor of $2$ comes from the two polarisation degrees of freedom of the electromagnetic field. Going now to the electric/magnetic field energy densities per logarithmic interval of $k$, the latter are defined as 
\beq\label{eq:EB_proper_vs_drhoEB_dlnk}
\frac{\mathrm{d}\rho_\mathrm{B}(k,\eta)}{\mathrm{d}\ln k}\equiv \frac{1}{2}\frac{4\pi k^3}{(2\pi)^3}|B^{\mathrm{proper}}(k,\eta)|^2f^2(\eta),\quad \frac{\mathrm{d}\rho_\mathrm{E}(k,\eta)}{\mathrm{d}\ln k}\equiv \frac{1}{2}\frac{4\pi k^3}{(2\pi)^3}|E^{\mathrm{proper}}(k,\eta)|^2f^2(\eta).
\eeq
and one can show straightforwardly that~\cite{Subramanian:2009fu}
\beq\label{eq:drho_EB_dlnk_definition}
\frac{\mathrm{d}\rho_\mathrm{B}(k,\eta)}{\mathrm{d}\ln k} = \frac{1}{2\pi^2}\frac{k^5}{a^4}|\mathcal{A}(k,\eta)|^2\quad\mathrm{and}\quad \frac{\mathrm{d}\rho_\mathrm{E}(k,\eta)}{\mathrm{d}\ln k} = \frac{f^2}{2\pi^2}\frac{k^3}{a^4} \left\vert \left[\frac{ \mathcal{A}(k,\eta)}{f}\right]^\prime\right\vert ^2.
\eeq

At this point, we should stress that in the following we will consider $\alpha = 2$, i.e. $\nu_1 = 4$, giving rise to a scale-invariant magnetic spectral density during inflation as it can be seen by plugging \Eq{eq:A_inf_superhorizon} into \Eq{eq:drho_EB_dlnk_definition}. This choice is justified since the magnetic field strength is not strongly peaked at a particular scale, hence a scale-invariant spectrum satisfies the currently available experimental data which give us only an upper bound on the values of the amplitude of the primordial magnetic field~\cite{Bamba:2003av,Martin:2007ue}.

At the end, fixing $\alpha = 2$ and plugging thus \Eq{A_reh_superhorizon} into \Eq{eq:drho_EB_dlnk_definition} and keeping the dominant terms, we obtain that the electric and magnetic energy densities per logarithmic comoving scale $k$ during the post-inflationary era on superhorizon scales read as
\begin{eqnarray}\label{eq:drho_E_dlnk_reh}
\frac{\mathrm{d}\rho_E(k,\eta)}{\mathrm{d}\ln k} & \simeq \frac{2.5\times 10^{-3}}{a^4}\left[1+ a^4_\mathrm{inf}\left(\frac{H_\mathrm{inf}}{\Mp}\right)^2\left(\frac{a}{a_\mathrm{inf}}\right)^{-2\beta}\right]\left(\frac{k}{a_\mathrm{inf}H_\mathrm{inf}}\right)^2H^4_\mathrm{inf}\left(\frac{a}{a_\mathrm{inf}}\right)^{4\beta}\\
\frac{\mathrm{d}\rho_B(k,\eta)}{\mathrm{d}\ln k} &  \simeq \frac{2.5\times 10^{-3}}{a^4}\left[1+ a^4_\mathrm{inf}\left(\frac{H_\mathrm{inf}}{\Mp}\right)^2\left(\frac{a}{a_\mathrm{inf}}\right)^{-2\beta}\right]\left(\frac{k}{a_\mathrm{inf}H_\mathrm{inf}}\right)^4H^4_\mathrm{inf}\left(\frac{a}{a_\mathrm{inf}}\right)^{4\beta+1} \label{eq:drho_B_dlnk_reh}
\end{eqnarray}
For modes which are on the other hand sub-horizon, i.e. $k\geq aH$, one can approximate their electric/magnetic spectral energy density by substituting $\eta=\eta_k$ at \Eq{eq:drho_E_dlnk_reh} and \Eq{eq:drho_B_dlnk_reh}~\cite{Sharma:2019jtb,Brandenburg:2021pdv,Brandenburg:2021bfx}, where $\eta_k$ is the conformal time when a mode $k$ crosses the cosmological horizon.

Finally, one can introduce the electric/magnetic field power spectrum $P_{E/B}$ which by definition is related to $\frac{\mathrm{d}\rho_{E/B}(k,\eta)}{\mathrm{d}\ln k}$ as
\beq\label{eq:def_P_B}
P_E(k,\eta) \equiv \frac{(2\pi)^3}{k^3}\frac{\mathrm{d}\rho_E(k,\eta)}{\mathrm{d}\ln k},\quad P_B(k,\eta) \equiv \frac{(2\pi)^3}{k^3}\frac{\mathrm{d}\rho_B(k,\eta)}{\mathrm{d}\ln k}.
\eeq
Note as well that $P_{B}$ and $P_{E}$ are connected to the equal time correlator of $E$ and $B$ respectively, which for the case of non-helical magnetic fields is defined as~\cite{Caprini:2009fx}
\begin{align*}
\langle E_{i}(k,\eta) E^*_{j}(k',\eta) \rangle & = (2 \pi)^3 \delta(\vec{k}-\vec{k'})(\delta_{ij}-\hat{k}_i \hat{k}_j)P_{E}(k,\eta) \\
\langle B_{i}(k,\eta) B^*_{j}(k',\eta) \rangle & = (2 \pi)^3 \delta(\vec{k}-\vec{k'})(\delta_{ij}-\hat{k}_i \hat{k}_j)P_{B}(k,\eta).
\end{align*}
The generalisation to helical EM fields is straightforward~\cite{Sharma:2019jtb}.

\subsection{Avoiding the strong coupling and the backreaction issues}\label{sec:strong_coupling_backreaction}
Having extracted before the electric and magnetic field power spectra, it is important to highlight here some issues present whenever the electromagnetic Lagrangian is coupled to an external field through a coupling of the form $f^2(R,\phi,...)F^{\mu\nu}F_{\mu\nu}$. In particular, the effective electric charge is defined as $e_\mathrm{eff} \equiv e/f^2$, hence whenever the coupling function $f$ takes very small values, this will inevitably lead to unacceptably strong couplings between charged particles and the EM field, making thus the perturbative treatment of an EM test field untrustworthy. 

In order to avoid such a problem, one needs to start with a value of $f$ at the end of inflation close to unity and then evolve it in such a way that it takes a large value at the end of inflation, so that the effective electric charge is small. Then, $f$ should decrease so as to acquire again its pre-inflationary value and restore the standard couplings at reheating. In order thus to recover classical electromagnetism at reheating, we need to have $f^2(a_\mathrm{reh})\simeq 1$, which for our case, where $\alpha =2$, combining \Eq{eq:f_definition} and \Eq{eq:xi_parametrisation} imposes the following constraint on the parameters at hand, namely $H_\mathrm{inf}$, $\Delta N_\mathrm{inf}$, $\Delta N_\mathrm{reh}$ and $\beta$:
\beq\label{eq:alpha_beta_constraints}
\xi(a_\mathrm{reh})R(a_\mathrm{reh})< 1 \Longrightarrow \beta \Delta N_\mathrm{reh} - 2\Delta N_\mathrm{inf} - \ln\left(\frac{H_\mathrm{inf}}{\Mp}\right)>0,
\eeq
where $\Delta N_\mathrm{inf}$ and $\Delta N_\mathrm{reh}$ are the e-folds elapsed during inflation and the post-inflationary era respectively and $H_\mathrm{inf}$ the Hubble scale at the end of inflation.

On the other hand, the back-reaction issue is related to the fact that the electric and magnetic energy densities should not overshoot the total background energy density throughout the cosmic evolution up to reheating, i.e.
\beq\label{eq:backreaction_constraint_general}
\rho_E + \rho_B < \rho_\mathrm{tot}.
\eeq

During inflation, it has been shown that for a scale-invariant magnetic field power spectrum, i.e. $\alpha=2$, there is no such problem~\cite{Sharma:2017eps}. In order to ensure that such back-reaction issue is also absent during the post-inflationary era, one should impose \Eq{eq:backreaction_constraint_general} for $\eta = \eta_\mathrm{reh}$. In particular, one can write \Eq{eq:backreaction_constraint_general} at reheating as
\beq\label{eq:backreaction_constraint_reheating}
\int_{k_\mathrm{ini}}^{k_\mathrm{reh}}\frac{\mathrm{d}\rho_E(k,\eta_\mathrm{reh})}{\mathrm{d}\ln k}\mathrm{d}\ln k + \int_{k_\mathrm{ini}}^{k_\mathrm{reh}}\frac{\mathrm{d}\rho_B(k,\eta_\mathrm{reh})}{\mathrm{d}\ln k}\mathrm{d}\ln k < g_\mathrm{reh}\frac{\pi^2}{30}T^4_\mathrm{reh},
\eeq
where $k_\mathrm{ini}$ and $k_\mathrm{reh}$ stand for the modes crossing the cosmological horizon at the onset of inflation and at the end of the post-inflationary era respectively, $g_\mathrm{reh}$ for the number of relativistic degrees of freedom at the end of reheating and $T_\mathrm{reh}$ for the reheating temperature. As it can be noted from \Eq{eq:backreaction_constraint_reheating}, the integral over $\ln k$ runs from $k_\mathrm{ini}$ to $k_\mathrm{reh}$, since at end of reheating only the modes between $k_\mathrm{ini}$ and $k_\mathrm{reh}$ are super-horizon and have been amplified according to \Eq{eq:drho_E_dlnk_reh} and \Eq{eq:drho_B_dlnk_reh}. At the end, accounting for the fact that $k_\mathrm{ini} = a_\mathrm{inf}H_\mathrm{inf}e^{-\Delta N_\mathrm{inf}}$ and $k_\mathrm{reh} =  a_\mathrm{inf}H_\mathrm{inf}e^{-\Delta N_\mathrm{reh}/2}$, one can straightforwardly obtain that the free parameters of our physical setup, namely $H_\mathrm{inf}$, $\Delta N_\mathrm{inf}$, $\Delta N_\mathrm{reh}$ and $\beta$, should fulfill the following constraint inequality:
\beq
\ln\left[\frac{30C}{g_\mathrm{reh}\pi^2}\left(\frac{1+2(2\beta + \frac{1}{2})^2}{4(2\beta + \frac{1}{2})^2}\right)\right] + 4\ln\left(\frac{H_\mathrm{inf}}{T_\mathrm{reh}}\right) <4\Delta N_\mathrm{inf} - (4\beta - 5)\Delta N_\mathrm{reh},
\eeq
where $C = 2.5\times 10^{-3}$.

\section{Electromagnetically induced gravitational waves}\label{sec:EMIGWs}

Let us now focus on the gravitational waves induced by the electromagnetic field. To do so, we shall consider the first order tensor perturbations with the perturbed Friedmann–Lema\^itre–Robertson–Walker (FLRW) background metric written as
$$\mathrm{d}s^2 = -\mathrm{d}t^2+a^2(t) \left( \delta_{ij} + 2h_{ij} \right)\mathrm{d}\vec{x}^2, $$ where $a$ is the scale factor and $h$ is the transverse ($\partial_i h^{ij} = 0$) and traceless ($h^i_i = 0$) first order tensor perturbation.

\subsection{The equation of motion for the tensor modes}
In order to extract the equation for the tensor modes, we will expand our field equations \eqref{eq: GravField equation} to first order in $h$. It will prove useful computationally to phrase our analysis in terms of the trace of the Faraday tensor $F \equiv F_{\alpha \beta} F^{\alpha \beta}$, which is a scalar quantity. In this context, after a long but straightforward calculation the ``space-space" component of our field equations read as
\bea\label{eq:tensor_perturbations_full_real_space}
& \frac{1}{\kappa^2}\left[\delta^i_j \left( -3 H^2 - 2\dot{H} \right) + \ddot{h}^i_j + 3H \dot{h}^i_j - \frac{\Delta}{a^2} h^i_j \right]=
 \left[1 + 6 \xi\left( \dot{H} + 2H^2 \right) \right] \Big[ -F^i_0 F_{j0} + \frac{1}{a^2} F^i_l F_{jl} - \\
& 2\frac{h^{lm}}{a^2} F^i_m F_{jl} - \frac{1}{4} \delta^i_j F  \Big] %+ 6 \xi_2 \left( \dot{H} + 2H^2 \right) \Big[ -F^i_0 \tilde{F}_{j0} + \frac{1}{a^2} F^i_l \tilde{F}_{jl} - 2\frac{h^{lm}}{a^2} F^i_m \tilde{F}_{jl} - \frac{1}{4} \delta^i_j \tilde{F}  \Big]+ \\ 
+\frac{ \xi }{2} \Big[ \delta^i_j F (\dot{H} + 3H^2) + F \Big( \ddot{h}^i_j + 3H \dot{h}^i_j - \frac{\Delta}{a^2} h^i_j \Big) + \delta^i_j \left( \Box F\right) - \frac{F_{,ij}}{a^2} \\ &  + \delta^ i_j H \dot{F} + 2 \frac{h^{li}}{a^2} F_{,jl} + \dot{h}_{ij} \dot{F}+  \frac{1}{a^2} \left( h_{im,j}+ h_{mj,i} - h_{ij,m} \right) F_{,m} \Big], %+ \xi_2 \frac{1}{2} \Big[ \delta^i_j \tilde{F} (\dot{H} + 3H^2) + \tilde{F} \Big( \ddot{h}^i_j + 3H \dot{h}^i_j - \frac{\Delta}{a^2} h^i_j \Big)+ \\ & \delta^i_j \left( \Box \tilde{F} \right) - \frac{\tilde{F}_{,ij}}{a^2} + \delta^ i_j H \dot{\tilde{F}} + 2 \frac{h^{li}}{a^2} \tilde{F}_{,jl} + \dot{h}_{ij} \dot{\tilde{F}}+  \frac{1}{a^2} \left( h_{im,j}+ h_{mj,i} - h_{ij,m} \right) \tilde{F}_{,m} \Big] , 
\eea
where $H \equiv \dot{a}/a $ is the Hubble function, $\Delta \equiv \partial^i \partial_i$ and comma denotes partial differentiation. Now to isolate the $h$ terms, we need to act with the following projection tensor $ \mathcal{O}_{ij,lm}= \mathcal{P}_{il}\mathcal{P}_{jm} - \frac{1}{2} \mathcal{P}_{ij}\mathcal{P}_{lm} $ with $\mathcal{P}_{ij} \equiv \delta_{ij} - \hat{k}_i\hat{k}_j$ and go to Fourier space via the transformation $h_{ij}(\vec{x},t) = \int \frac{d^3k}{(2\pi)^3}h_{ij}(\vec{k},t) e^{i \vec{k} \cdot \vec{x}}  $. After this operation, all the terms with the $\delta^i_j$ will be eliminated. Concerning the rest of the terms, we shall make the following approximations: Since we treat the vector gauge field as a background quantity, it should respect the symmetry of the background metric, namely the FLRW metric, which is homogeneous and isotropic. Thus, one expects a mild dependence of $A_\mu$ on the spatial coordinates~\cite{Bamba:2008ja}. Consequently, terms in \Eq{eq:tensor_perturbations_full_real_space} which contain two spatial derivatives of the vector potential, like $\partial_k\partial_l A_\mu$, and $(\partial_k A_\mu)^2 h_{ij}$ can be neglected, leaving us with  

\beq
\begin{split}\label{eq:h_eq_of_motion_real_space}
& h^{\prime\prime}_{ij}\left(1 - \frac{\xi F}{2\Mp^4}\right) + 3\mathcal{H}\left[1 - \frac{\xi}{2\Mp^4}\left(F  + \frac{F^\prime}{3\mathcal{H}}\right)\right]{h}^\prime_{ij} + \left(1- \frac{\xi F}{2\Mp^4}\right) \Delta h_{ij}  = \\
& \frac{1}{\Mp^2} \left[1 + 6 \frac{\xi}{\Mp^2}\left( \mathcal{H}^\prime + 2\mathcal{H}^2 \right)\right] \mathcal{O}_{ij,lm}\Big( -F^l_0 F_{m0} + \frac{1}{a^2} F^l_p F_{mp}\Big),
\end{split}
\eeq
where we have replaces $\kappa^2$ with $1/\Mp^2$ and prime denotes differentiation with respect to the conformal time $\eta$, defined as $dt \equiv a d\eta$.

At this point, it helps our intuition if we work explicitly with the electric and magnetic fields, which are related to the Faraday tensor as
\beq
\begin{split}
& E_i = \frac{F_{0i}}{a} = -\frac{1}{a} A'_i  \\
& B_i = \frac{1}{2a} \epsilon_{ijk} \delta^{jl} \delta^{km} F_{lm}= \frac{1}{a} \epsilon_{ijk} \delta^{jl} \delta^{km} \partial_l A_m. \label{EB}
\end{split}
\eeq
Since $F \sim E^2 + B^2 + E \cdot  B $, the terms $Fh$, which contain spatial derivatives of the gauge field, namely the $B^2h$ and $E \cdot  B h$ terms, can be safely neglected following the same reasoning described above. Hence we keep only terms proportional to $E^2$. At the end, one is met with the following equation of motion for the Fourier modes of the tensor perturbations:

\beq\label{eq:h_eq_of_motion_k_space}
\begin{split}
& h''_{\boldmathsymbol{k}} - \frac{\xi}{2\Mp^4} \int h''_{\boldmathsymbol{k'}}F_{\boldmathsymbol{k'-k}}\mathrm{d}\boldmathsymbol{k}'   +  2\calH h'_\boldmathsymbol{k} - \frac{3\calH \xi}{2\Mp^4}\int h'_{\boldmathsymbol{k'}}F_{\boldmathsymbol{k'-k}}\mathrm{d}\boldmathsymbol{k}' - \frac{\xi}{2\Mp^4}\int h'_{\boldmathsymbol{k'}}F^\prime_{\boldmathsymbol{k'-k}}\mathrm{d}\boldmathsymbol{k}'
\\ & + k^2 \left(h_\boldmathsymbol{k} - \frac{\xi}{2 \Mp^4} \int h_{\boldmathsymbol{k'}}F_{\boldmathsymbol{k'-k}}\mathrm{d}\boldmathsymbol{k}' \right)  = \frac{1}{\Mp^2} \left[1 + \frac{6\xi}{\Mp^2} \left( \mathcal{H}^\prime+ \mathcal{H}^2 \right)\right] \Pi_\boldmathsymbol{k},
 \end{split}
\eeq
where we have used the property of the Fourier transforms that the Fourier transform of the product of two functions is the convolution of their Fourier transforms. 

In \Eq{eq:h_eq_of_motion_k_space}, $\Pi_\boldmathsymbol{k}$ is the Fourier transform of the anisotropic electromagnetic stress which is defined as $\Pi_{ij} \equiv  \mathcal{O}_{ij,lm} \left(-F^l_0 F_{m0} + \frac{1}{a^2} F^l_p F_{mp} \right)$. Note here that it can be straightforwardly shown [See Appendix \ref{app:The trace F}] that ignoring $B^2$ and $E \cdot  B $ terms,  $F =  - 2 E^2_{\mathrm{proper}}(\eta,\boldmathsymbol{x})$ while in Fourier space $F = - 2 E^2_{\mathrm{proper}}(k,\eta)$ with $E^2_{\mathrm{proper}}(k,\eta)$ given by \Eq{eq:EB_proper_vs_drhoEB_dlnk}. As it was checked numerically [See Appendix \ref{app:The coupling terms}] for the choice of parameters $\Delta N_\mathrm{inf}$, $\Delta N_\mathrm{reh}$, $H_\mathrm{inf}$ and $\beta$ we chose to work with, the terms $\frac{\xi F}{2\Mp^4}$ and $ \frac{F^\prime}{6 \calH F \Mp^4 }$ modifying the tensor mode dispersion relation are orders of magnitude smaller than unity, thus they can be safely neglected. Thus, \Eq{eq:h_eq_of_motion_k_space} becomes
\beq\label{eq:h_k_dispersion_k_space}
h''_{\boldmathsymbol{k}}   +  2\calH h'_\boldmathsymbol{k} + k^2 h_\boldmathsymbol{k} = \frac{1}{\Mp^2} \left[1 + \frac{6\xi}{\Mp^2} \left( \mathcal{H}^\prime+ \mathcal{H}^2 \right)\right] \Pi_\boldmathsymbol{k}
\eeq

We need to emphasize here that in general the terms $\frac{\xi F}{2\Mp^4}$ and $ \frac{F^\prime}{6 \calH F \Mp^4 }$ can be larger than unity. In this case one needs to solve fully numerically the integral-differential equation \eqref{eq:h_eq_of_motion_k_space} in $k$ space accounting for the convolution integrals coupling the tensor field with the Faraday tensor, something which may necessitate the use of renormalisation techniques of quantum field theory~\cite{Mazzitelli:1995mp}, which goes beyond the scope of this work and shall be investigated elsewhere. Thus, what is needed now to proceed to the GW spectrum derivation, is the explicit relation of $\Pi_\boldmathsymbol{k}$ to the electric and magnetic field power spectrum in $k$ space. This is the subject of the following section.
%%%%%%%%%%%%%%%%%%%%
\subsection{The anisotropic electromagnetic stress} 
Based on the analysis of \cite{Sharma:2019jtb}, the relevant terms of the anisotropic electromagnetic stress $\Pi_{ij}$ in Fourier space in terms of the electric and magnetic fields are given by
\beq
\Pi_{ij} (\vec{k}, \eta) = \frac{1}{4 \pi a^2} \int \frac{d^3q}{2 \pi^3}  \mathcal{O}_{ij}^{lm} \Big( B_m (\vec{q}, \eta) B^{\star}_n(\vec{q}-\vec{k},\eta)  + E_m (\vec{q}, \eta) E^{\star}_n(\vec{q}-\vec{k},\eta)\Big).
\eeq

We can then define the equal-time 2-point correlator of the anisotropic electric and magnetic stresses as:
\beq
\langle \Pi^{E}_{ij}(\boldmathsymbol{k},\eta)\Pi^E_{ij}(\boldmathsymbol{q},\eta)\rangle    \equiv \Pi_E(k,\eta)\delta(\boldmathsymbol{k}-\boldmathsymbol{q}),\quad \langle \Pi^B_{ij}(\boldmathsymbol{k},\eta)\Pi^B_{ij}(\boldmathsymbol{q},\eta)\rangle    \equiv \Pi_B(k,\eta)\delta(\boldmathsymbol{k}-\boldmathsymbol{q}),
\eeq
where $\Pi_E(\boldmathsymbol{k},\eta)$ and $\Pi_B(\boldmathsymbol{k},\eta)$ are the power spectrum of the anisotropic electric and the magnetic stresses respectively related to the electric and magnetic field power spectra $P_E(k,\eta)$ and $P_B(k,\eta)$ as follows~\cite{Caprini:2006jb}:
\beq\label{eq:Pi_B_vs_P_B}
\Pi_{E/B}(k,\eta) = \int \mathrm{d}^3\boldmathsymbol{q} P_{E/B}(q,\eta)P_{E/B}(|\boldmathsymbol{q}-\boldmathsymbol{k}|,\eta) (1+\gamma^2) (1+\beta^2),
\eeq
where $\gamma = \hat{\boldmathsymbol{k}}\cdot \hat{\boldmathsymbol{q}}$ and $\beta = \hat{\boldmathsymbol{k}}\cdot \widehat{\boldmathsymbol{k} - \boldmathsymbol{q}}$. 

Introducing now the auxiliary variables $v$ and $u$ such as that $u=|\boldmathsymbol{q}-\boldmathsymbol{k}|/k$ and $v=q/k$ after some algebraic manipulations one can recast $\Pi_{E/B}(k,\eta)$ in the following form:
\beq
\label{eq:Pi_B_vs_P_B_u_v}
\begin{split}
& \Pi_{E/B}(k,\eta) = 2\pi k^3 \int_{0}^\infty \mathrm{d}v \int_{|1-v|}^{1+v} \mathrm{d} u P_{E/B}(kv,\eta)P_{E/B}(ku,\eta) uv \\  & \times
\left[1+\frac{(1+v^2-u^2)^2}{4v^2}\right]\left[1+\frac{(1+u^2-v^2)^2}{4u^2}\right].
\end{split}
\eeq
%
%%%%%%%%%%%%%%%%%%%%%%%%%%%%
\subsection{The tensor mode dynamics}
Having extracted above the power spectrum of the anisotropic electric and magnetic stresses we study here the dynamics of the tensor perturbations $h_\boldmathsymbol{k}$ induced by such anisotropic stresses. In particular, the equation of motion for $h_\boldmathsymbol{k}$, namely \Eq{eq:h_k_dispersion_k_space}, can be recast as~\cite{Caprini:2006jb}:

\beq\label{eq:tensor_eq_motion}
h_\boldmathsymbol{k}^{s,\prime\prime} +  2\calH h_\boldmathsymbol{k}^{s,\prime} +  k^2h^s_\boldmathsymbol{k} =  \frac{1}{\Mp^2 a^2} \left[1 + \frac{6 \xi(\eta)}{\Mp^2} \left( \calH' + \calH^2 \right)  \right]\sqrt{\Pi_{E/B}(k,\eta)},
\eeq
where $s = (+), (\times)$ stands for the two polarisation states of the tensor modes in general relativity. 

An analytic solution to \Eq{eq:tensor_eq_motion} can be obtained by using Green's function formalism. Namely, one gets that 
\bea
\label{tensor mode function}
h^s_\boldmathsymbol{k} (\eta)  = \frac{1}{\Mp^2}  \int^{\eta}_{\eta_0}\mathrm{d}\bar{\eta}\, \left[1 +\frac{6 \xi(\eta)}{\Mp^2} \left( \calH' + \calH^2 \right)  \right]  G_\boldmathsymbol{k}(\eta,\bar{\eta})\frac{\sqrt{\Pi_B(k,\bar{\eta})}}{a^2(\bar{\eta})},
\eea
where $G_\boldmathsymbol{k}(\eta,\bar{\eta})$ is the Green's function, which is the solution of the equation \Eq{eq:tensor_eq_motion} with the source term replaced by a delta function, reading as 
\beq
\label{Green function equation}
G_\boldmathsymbol{k}^{s,\prime\prime}(\eta,\bar{\eta})  + 2\calH  G_\boldmathsymbol{k}^{s,\prime}(\eta,\bar{\eta}) +  k^{2}G^s_\boldmathsymbol{k}(\eta,\bar{\eta}) = \delta (\eta - \bar{\eta}),
\eeq
with the boundary conditions $\lim_{\eta\to \bar{\eta}}G^s_\boldmathsymbol{k}(\eta,\bar{\eta}) = 0$ and $ \lim_{\eta\to \bar{\eta}}G^{s,\prime}_\boldmathsymbol{k}(\eta,\bar{\eta})=1$. 
Using the scale factor parametrisation during the post-inflationary era, introduced in \Eq{eq:a_parametrisation}, one obtains that 
\beq
\label{Green function equation}
G_\boldmathsymbol{k}^{s,\prime\prime}(\eta,\bar{\eta})  + \frac{4e^{\Delta N_\mathrm{inf}}H_\mathrm{inf}}{3+ e^{\Delta N_\mathrm{inf}}H_\mathrm{inf}\eta} G_\boldmathsymbol{k}^{s,\prime}(\eta,\bar{\eta}) +  k^{2}G^s_\boldmathsymbol{k}(\eta,\bar{\eta}) = \delta (\eta - \bar{\eta}).
\eeq
Imposing then the boundary conditions $\lim_{\eta\to \bar{\eta}}G^s_\boldmathsymbol{k}(\eta,\bar{\eta}) = 0$ and $ \lim_{\eta\to \bar{\eta}}G^{s,\prime}_\boldmathsymbol{k}(\eta,\bar{\eta})=1$ one gets an analytic solution for the Green function being recast as
\beq\label{eq:G_k_reheating_MD}
\begin{split}
G^s_\boldmathsymbol{k}(\eta,\bar{\eta}) & = \frac{e^{-\Delta N_\mathrm{inf}} (3 + e^{\Delta N_\mathrm{inf}}H_\mathrm{inf}\bar{\eta})^{3/2}}{H_\mathrm{inf} k^3 \sqrt{\frac{3e^{-\Delta N_\mathrm{inf}}}{H_\mathrm{inf}} + \eta} \left(3 + e^{\Delta N_\mathrm{inf}}H_\mathrm{inf} \eta\right)^{5/2} \sqrt{\frac{3e^{-\Delta N_\mathrm{inf}}}{H_\mathrm{inf}} + \bar{\eta}}} \\ & 
\times \Biggl[-e^{2\Delta N_\mathrm{inf}}H^2_\mathrm{inf} k(\eta - \bar{\eta}) \cos\left(k (\eta - \bar{\eta})\right) \\ & 
+ \Bigl\{9 k^2 + e^{\Delta N_\mathrm{inf}}H_\mathrm{inf} \left[3 k^2 (\eta + \bar{\eta}) +  e^{\Delta N_\mathrm{inf}}H_\mathrm{inf} \left(1 + k^2 \eta\bar{\eta}\right)\right]\Bigr\} \sin\left(k(\eta - \bar{\eta})\right)\Biggr]
\end{split}
\eeq

%\beq\label{eq:G_k_MD}
%kG^{\mathrm{MD}}_\boldmathsymbol{k}(\eta,\bar{\eta})  = \frac{1}{x\bar{x}}\left[(1+x\bar{x})\sin\left(x-\bar{x}\right)-(x-\bar{x})\cos\left(x-\bar{x}\right)\right],
%\eeq
%where $x\equiv k\eta$.

\subsection{The electromagnetically induced gravitational-wave signal}
One then can define the GW spectral abundance as $\OmegaGW (\eta,k)\equiv \frac{1}{\rho_\mathrm{c}}\frac{\mathrm{d}\rhoGW}{\mathrm{d}\ln k}$, where $\rho_\mathrm{c}=3H^2/(8\pi G)$ is the critical energy density, and show that, in the case of free propagating tensor modes,  $\OmegaGW (\eta,k)$  can be recast as~\cite{Maggiore:1999vm} 
\bea
\label{Omega_GW_eta_k}
\OmegaGW (\eta,k) = \frac{1}{24}\left[\frac{k}{\calH(\eta)}\right]^{2}\overline{\mathcal{P}}_{h}(\eta,k), 
\eea
where $\calH$ is the conformal Hubble parameter and $\mathcal{P}_{h}(\eta,k),$ is the tensor power spectrum defined as 
\beq\label{eq:P_h}
\mathcal{P}_{h}(\eta,k) \equiv \frac{k^3|h_\boldmathsymbol{k}|^2}{2\pi^2}.
\eeq
The bar denotes averaging over the sub-horizon oscillations of the tensor field, which is done in order to extract the envelope of the GW spectrum at those scales. 

Finally, accounting for the fact that GW radiation scales as $a^{-4}$ due to cosmic expansion and considering that the 
radiation energy 
density reads as $\rho_r = 
\frac{\pi^2}{30}g_{*\mathrm{\rho}}T_\mathrm{r}^4$, where the primordial plasma temperature $T_\mathrm{r}$ scales as $T_\mathrm{r}\propto 
g^{-1/3}_{*\mathrm{S}}a^{-1}$, one obtains that $\OmegaGW (\eta,k)$ 
today reads as
\beq\label{Omega_GW_RD_0}
\Omega_\mathrm{GW}(\eta_0,k) = 
\Omega^{(0)}_r\frac{g_{*\mathrm{\rho},\mathrm{*}}}{g_{*\mathrm{\rho},0}}
\left(\frac{g_{*\mathrm{S},\mathrm{0}}}{g_{*\mathrm{S},\mathrm{*}}}\right)^{4/3}
\OmegaGW(\eta_\mathrm{*},k),
\eeq
where $g_{*\mathrm{\rho}}$ and $g_{*\mathrm{S}}$ stand for the energy and entropy relativistic degrees of freedom.

\begin{figure*}[h!]
\begin{center}
\includegraphics[width=0.796\textwidth]{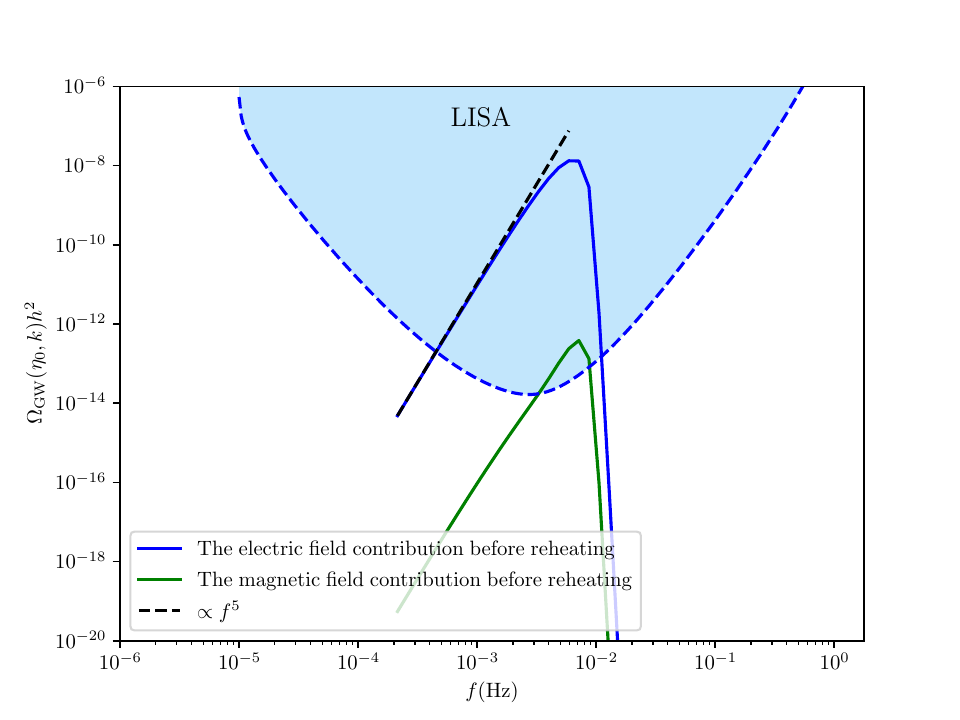}
\caption{{\it{The gravitational-wave signal induced by gravity-electromagnetism coupling of the form $RF^{\mu\nu}F_{\mu\nu}$ before reheating for $H_\mathrm{inf} = 10^{-10}\Mp$, $\Delta N_\mathrm{inf}=45$, $\Delta N_\mathrm{reh}=25$ and $\beta=5.58$. In the blue curve we show the electric field contribution to the aforementioned GW signal before reheating while in the green one we depict the magnetic field contribution. In the dashed black line, we show the universal IR $f^5$ frequency scaling while on top of the electromagnetically induced GW signals we superimpose the sensitivity curve of  LISA~\cite{LISACosmologyWorkingGroup:2022jok}. }}}
\label{fig:Omega_GW_E_vs_B}
\end{center}
\end{figure*}

Note that the reference conformal time $\eta_\mathrm{*}$ in our case is considered to be the end of the post-inflationary era $\eta_\mathrm{reh}$, after which we are met with a thermalised primordial plasma with a very high conductivity, destroying the main source of our GW signal, namely the electric field. One then after $\eta_\mathrm{reh}$ can safely treat the EMIGWs as freely propagating tensor modes and use \Eq{Omega_GW_eta_k} to derive $\OmegaGW (\eta,k)$. Regarding the GWs induced by the magnetic field power spectrum during the post-inflationary era, these are subdominant compared to the ones induced by the electric field as it can be seen from \Fig{fig:Omega_GW_E_vs_B}. In principle, one should have also considered however the GWs induced by the magnetic field power spectrum after reheating. In order to derive such a GW signal, we should have performed computationally high cost magnetohydrodynamic (MHD) numerical simulations~\cite{Brandenburg:2021pdv,Brandenburg:2021bfx}. Nevertheless, according to recent analytic approximations~\cite{Sharma:2019jtb}, the magnetic contribution from the post-reheating era to the induced GW signal peaks at the same frequency as the electrically induced GW signal produced before reheating while the respective GW amplitudes differ less than one order of magnitude at the peak frequency, with the dominant contribution coming from the electrically induced GWs considered here.

\begin{figure*}[h!]
\begin{center}
\includegraphics[width=0.796\textwidth]{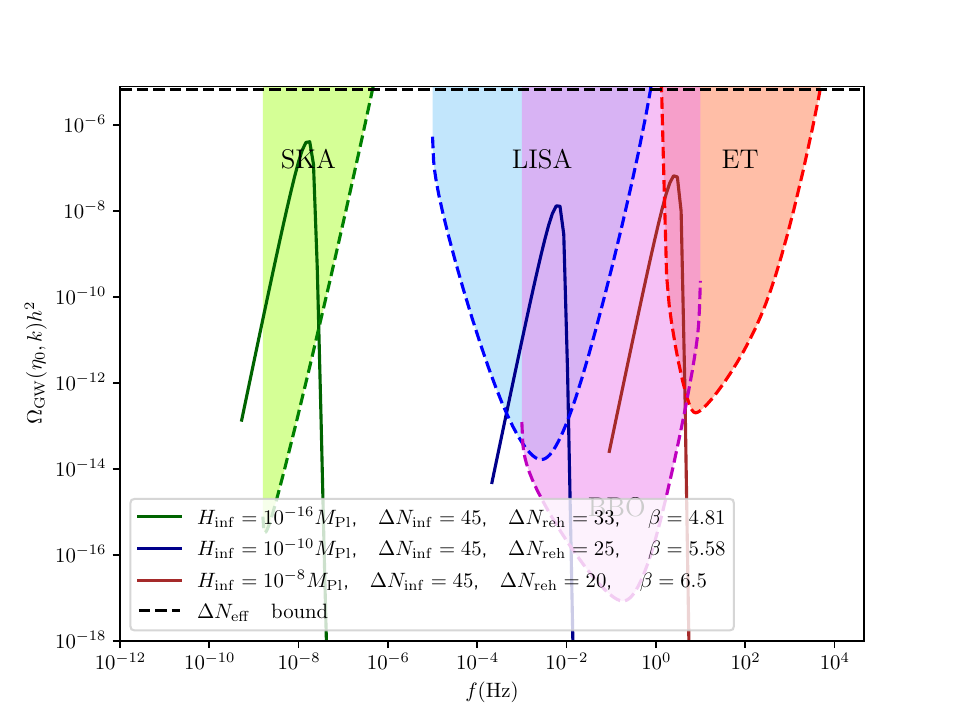}
\caption{{\it{The gravitational-wave signal induced from gravity-electromagnetism coupling of the form $RF^{\mu\nu}F_{\mu\nu}$ for different values of the parameters $H_\mathrm{inf}$, $\Delta N_\mathrm{inf}$, $\Delta N_\mathrm{reh}$ and $\beta$. On top of the  GW spectra we present  the sensitivity curves of SKA~\cite{Janssen:2014dka}, LISA~\cite{LISACosmologyWorkingGroup:2022jok}, BBO~\cite{Harry:2006fi} and ET~\cite{Maggiore:2019uih} GW experiments. In the horizontal black dashed line, we show also the upper bound on $\Omega_\mathrm{GW,0}h^2\leq 6.9\times 10^{-6}$ coming from the upper bound constraint on $\Delta N_\mathrm{eff}$ from CMB and BBN observations~\cite{Smith:2006nka}.}}}
\label{fig:Omega_GW}
\end{center}
\end{figure*}

In \Fig{fig:Omega_GW} we show the GW signal induced by  gravito-electromagnetic couplings of the form $RF^{\mu\nu}F_{\mu\nu}$ for different values of our parameters at hand, namely the energy scale at the end of inflation, $H_\mathrm{inf}$, the duration of inflation and reheating in e-folds, $\Delta N_\mathrm{inf}$ and $\Delta N_\mathrm{reh}$, and the exponent $\beta$, determining the dynamical behaviour of the gauge coupling function during the post-inflationary era. On top of the GW spectra in \Fig{fig:Omega_GW}  we superimpose the detection sensitivity curves of the GW experiments SKA~\cite{Janssen:2014dka}, LISA~\cite{LISACosmologyWorkingGroup:2022jok}, BBO~\cite{Harry:2006fi} and ET~\cite{Maggiore:2019uih}. 

Regarding the choice of our parameters at hand we need to mention that for the duration of inflation in e-folds we chose a fiducial value of $\Delta N_\mathrm{inf} = 45$, motivated by the Planck constraints on the duration of the inflation epoch~\cite{Planck:2018jri}. The other three parameters, namely $\Delta N_\mathrm{reh}$, $H_\mathrm{inf}$ and $\beta$ are constrained only by the strong coupling and backcreaction considerations discussed in \Sec{sec:strong_coupling_backreaction}.

Interestingly enough, we found a universal $f^5$ infrared (IR) frequency scaling [See \Fig{fig:Omega_GW_E_vs_B}] with the GW amplitude being well within the sensitivity detection bands of future GW experiments, hence being potentially detectable in the future in GW observatories serving in this way as a novel probe of gravito-electromagnetic couplings and inflationary magnetogenesis. 

\newpage
\section{Conclusions}\label{sec:Conclusions}
Gravitational waves can act undoubtedly as messengers from different epochs throughout the cosmic history. Notably, they can give us access to the early Universe shedding light to the physics of inflation, primordial phase transitions, topological defects and primordial black holes. 

In this paper, we focused on the portal of GWs induced by electromagnetic fields during the post-inflationary epoch through theoretically motivated non-minimal couplings of the gravity sector with electromagnetism on curved spacetime breaking the conformal invariance. In particular, we studied here the simplest non-minimal coupling term of the form $\xi RF^{\mu\nu}F_{\mu\nu}/\Mp^2$, where $\xi$ is a time-dependent gauge coupling function. Our formalism can be straightforwardly generalised to higher-order terms as well of the form $\mathcal{R}^nF^{\mu\nu}F_{\mu\nu}/\Mp^{2n}$, where $\mathcal{R}$ is any component of the Riemann tensor and $n$ a free parameter.

However, it is important to emphasize that such conformally breaking coupling terms contribute also to the generation of primordial magnetic fields, suffering however from strong coupling and back-reaction issues. In this paper, we accounted for these issues and set constraints on the free parameters of our physical setup, namely the duration in e-folds of inflation and the post-inflation epoch before reheating, $\Delta N_\mathrm{inf}$ and $\Delta N_\mathrm{reh}$, the energy scale at the end of inflation $H_\mathrm{inf}$ as well the exponent $\beta$ governing the time-dependence of the gauge coupling function before reheating.

Remarkably, after deriving the modified tensor mode dispersion relation \eqref{eq:h_eq_of_motion_k_space}, we found that coupling terms of the form $\mathcal{R}^nF^{\mu\nu}F_{\mu\nu}/\Mp^{2n}$ can modify both the propagation and the source of these electromagnetically induced GWs. In the present study, we restricted ourselves to studying mainly the effect of curvature-electromagnetic couplings on the source of the EMIGWs. The most interesting effect is however at the level of the propagation of GWs, which requires to solve fully numerically the modified tensor mode equation of motion \eqref{eq:h_eq_of_motion_k_space}, which actually becomes an integral-differential equation demanding high-cost numerical simulations and/or advanced theoretical techniques to be solved, being beyond the scope of the present work.

\begin{figure*}[h!]
\begin{center}
\includegraphics[width=0.796\textwidth]{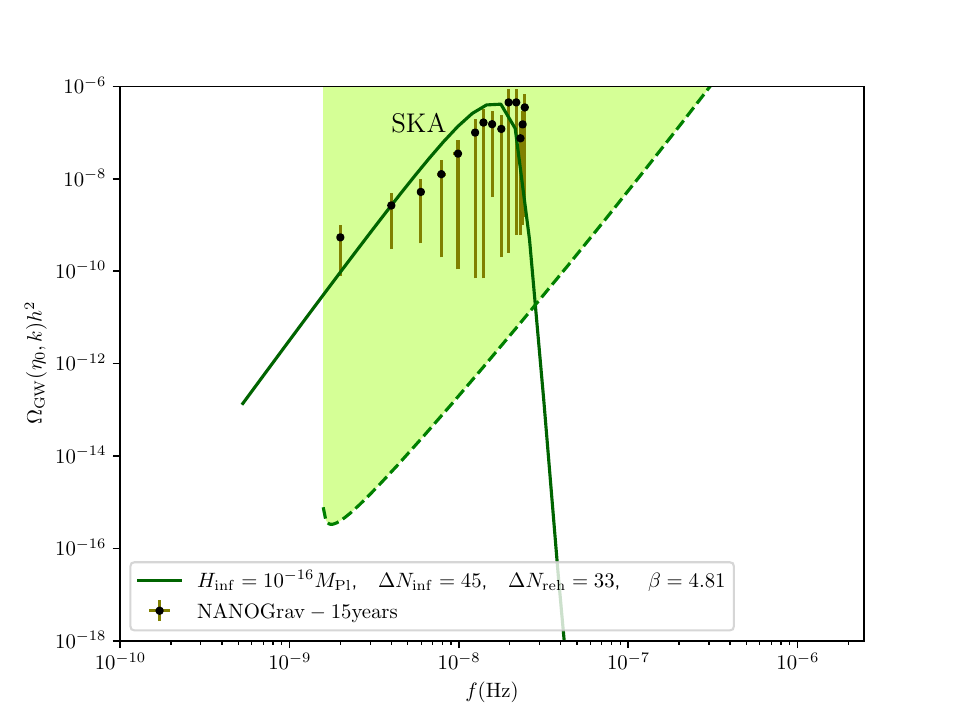}
\caption{{\it{The gravitational-wave signal induced by gravitional-electromagnetic coupling of the form $RF^{\mu\nu}F_{\mu\nu}$ before reheating for $H_\mathrm{inf} = 10^{-16}\Mp$, $\Delta N_\mathrm{inf}=45$, $\Delta N_\mathrm{reh}=33$ and $\beta=4.81$.  On top of the electromagnetically induced GW signal we superimpose the sensitivity curve of  SKA~\cite{Janssen:2014dka} as well the GW data released by the NANOGrav collaboration~\cite{NANOGrav:2023gor}.}}}
\label{fig:Omega_GW_NANOGrav}
\end{center}
\end{figure*}

Finally, we found that the dominant contribution to the electromagnetically induced gravitational wave signal produced before the onset of the radiation-dominated era mainly arises from the electric field and presents a universal infrared frequency scaling of $f^5$. This signal can fall well within the frequency detection bands of GW experiments, such as PTAs/SKA, LISA, ET, and BBO, making it potentially detectable by future GW observatories. What's more, as one can see from \Fig{fig:Omega_GW_NANOGrav}, for the fiducial choice of our free parameters $H_\mathrm{inf} = 10^{-16}\Mp$, $\Delta N_\mathrm{inf}=45$, $\Delta N_\mathrm{reh}=33$ and $\beta=4.81$, we obtain an EMIGW signal in quite good agreement with the recently released by NANOGrav $\mathrm{nHz}$ GW data~\cite{NANOGrav:2023gor}, signalling the GWs induced by gravito-electromagnetic couplings as a good candidate for the $\mathrm{nHz}$ GW background.

Our analysis can be extended in various research directions. In particular, one can study as well the EMIGW signal produced after reheating by performing general-relativistic MHD simulations during the radiation-dominated era, something which will mainly affect the spectral shape of the EMIGW signal. In addition, one can study as well the effect of modified gravity theories on the aforementioned GW signal both at the level of propagation and the GW source.

\vspace{0.5 cm}
{\bf Acknowledgements} -- 
TP and SC acknowledge the support of the INFN Sezione di Napoli \textit{initiativa specifica} QGSKY. All the authors recognise the contribution of the COST Action CA21136 ``Addressing observational tensions in cosmology with systematics and fundamental physics (CosmoVerse)''. In particular, a major part of this work was carried out during the stay of CT at Scuola Superiore Meriodionale which was facilitated thanks to his award of a Short Term Scientific Mission (STSM) grant from Cosmoverse. Further, TP and CT acknowledge the contribution of the LISA Cosmology Working Group and, finally, TP acknowledges financial support from the Foundation for Education and European Culture in Greece and CT from A.G. Leventis Foundation.

\appendix

\section{The trace $F=F^{\mu\nu}F_{\mu\nu}$ of the Faraday tensor}\label{app:The trace F}

We derive in this Appendix the trace $F=F^{\mu\nu}F_{\mu\nu}$ of the Faraday tensor. By definition, $F$ will be written in real space as
\beq
F=F^{\mu\nu}F_{\mu\nu} = 2F^{0i}F_{0i} + F^{ij}F_{ij} \simeq  -2\frac{\delta^{ij}}{a^2}\dot{A}_i\dot{A}_j = -2\frac{E^2(\boldmathsymbol{x},t)}{a^2}= - 2E^2_\mathrm{proper}(\boldmathsymbol{x},\eta),
\eeq
where we neglected the $F^{ij}F_{ij}$ terms being proportional to $(\partial_iA_j)^2$ given the mild spatial dependence of $A_\mu$ as mentioned in \Sec{sec:EMIGWs} and accounting for the fact that $E_i = -\dot{A}_i$. We also expressed in the last equality the time variable in terms of the conformal time.

In order now to get $E^2_\mathrm{proper}(\boldmathsymbol{x},\eta)$, one can apply an inverse Fourier transform on $E^2_\mathrm{proper}(k,\eta)$ given by \Eq{eq:EB_proper_vs_drhoEB_dlnk}. Doing so, we have that
\beq\label{E_proper_x}
E^2_\mathrm{proper}(\boldmathsymbol{x},\eta)\equiv \int_0^{k_\mathrm{inf}}\frac{\mathrm{d}^3\boldmathsymbol{k}}{(2\pi)^3}e^{i\boldmathsymbol{k}\cdot\boldmathsymbol{x}}E^2_\mathrm{proper}(k,\eta),
\eeq
with $E^2_\mathrm{proper}(k,\eta)$ being recast from \Eq{eq:EB_proper_vs_drhoEB_dlnk} as
\beq\label{E_proper_k}
E^2_\mathrm{proper}(k,\eta) = \frac{1}{f^2(\eta)}\frac{(2\pi)^3}{2\pi}\frac{1}{k^3}\frac{\mathrm{d}\rho_\mathrm{E}(k,\eta)}{\mathrm{d}\ln k}
\eeq
Using then spherical coordinates and plugging \Eq{E_proper_k} into \Eq{E_proper_x} one gets that
\beq
E^2_\mathrm{proper}(\boldmathsymbol{x},\eta) = 12.8\times 256^{-\beta}\frac{e^{-2\Delta N_\mathrm{inf}}H^2_\mathrm{inf}\left( 3+ e^{\Delta N_\mathrm{inf}}H_\mathrm{inf}x\right)^{8\beta - 8}}{r^2}\left[1 - \cos\left(e^{\Delta N_\mathrm{inf}}H_\mathrm{inf}r\right)\right].
\eeq

\section{The terms $\frac{\xi F}{2\Mp^4}$ and $\frac{\xi F^\prime}{6\mathcal{H}\Mp^4}$}\label{app:The coupling terms}

In order to give an order of magnitude estimate of the $\frac{\xi F}{2\Mp^4}$ and $\frac{\xi F^\prime}{6\mathcal{H}\Mp^4}$ terms present in the equation of motion of the tensor modes which modify the standard GR tensor perturbation dynamics, we will approximate $r$ as $r = \frac{2\pi}{k}a$. One then obtains straightforwardly that 
\beq
\begin{split}
\frac{\xi F}{2\Mp^4} = & 0.08\times 16^{-\beta}\frac{e^{-4\Delta N_\mathrm{inf}}H^2_\mathrm{inf}k^2\left( 3+ e^{\Delta N_\mathrm{inf}}H_\mathrm{inf}\eta \right)^{4\beta - 6}}{2\Mp^4}\\ & \times\left\{1  - \cos\left[\frac{\pi}{2k}e^{2\Delta N_\mathrm{inf}}H_\mathrm{inf}\left(3+e^{\Delta N_\mathrm{inf}}H_\mathrm{inf}\eta\right)^2\right]\right\},
\end{split}
\eeq

\beq
\begin{split}
\frac{\xi F^\prime}{6\mathcal{H}\Mp^4} & =  \frac{16^{-\beta}e^{-4\Delta N_\mathrm{inf}}H^2_\mathrm{inf}}{k}\frac{\left( 3+ e^{\Delta N_\mathrm{inf}}H_\mathrm{inf}\eta\right)^{4\beta - 6}}{2\Mp^4}\Biggl\{\left(-0.1621 + 0.108\beta\right)k \\ & + \left(-0.1621 - 0.108\beta\right)k\cos\left(\frac{\pi}{2k}e^{2\Delta N_\mathrm{inf}}H_\mathrm{inf}\left(3+e^{\Delta N_\mathrm{inf}}H_\mathrm{inf}\eta\right)^2\right) \\ & +   e^{2\Delta N_\mathrm{inf}}H_\mathrm{inf}\left[0.382 + e^{\Delta N_\mathrm{inf}}H_\mathrm{inf}\eta\left(0.254 + 0.042e^{\Delta N_\mathrm{inf}}H_\mathrm{inf}\eta\right)\right] \\ & \times \sin\left(\frac{\pi}{2k}e^{2\Delta N_\mathrm{inf}}H_\mathrm{inf}\left(3+e^{\Delta N_\mathrm{inf}}H_\mathrm{inf}\eta\right)^2\right)\Biggr\}.
\end{split}
\eeq

\begin{figure}[h!]
\centering
%\subfloat[]
{\label{a}\includegraphics[width=.5\linewidth]{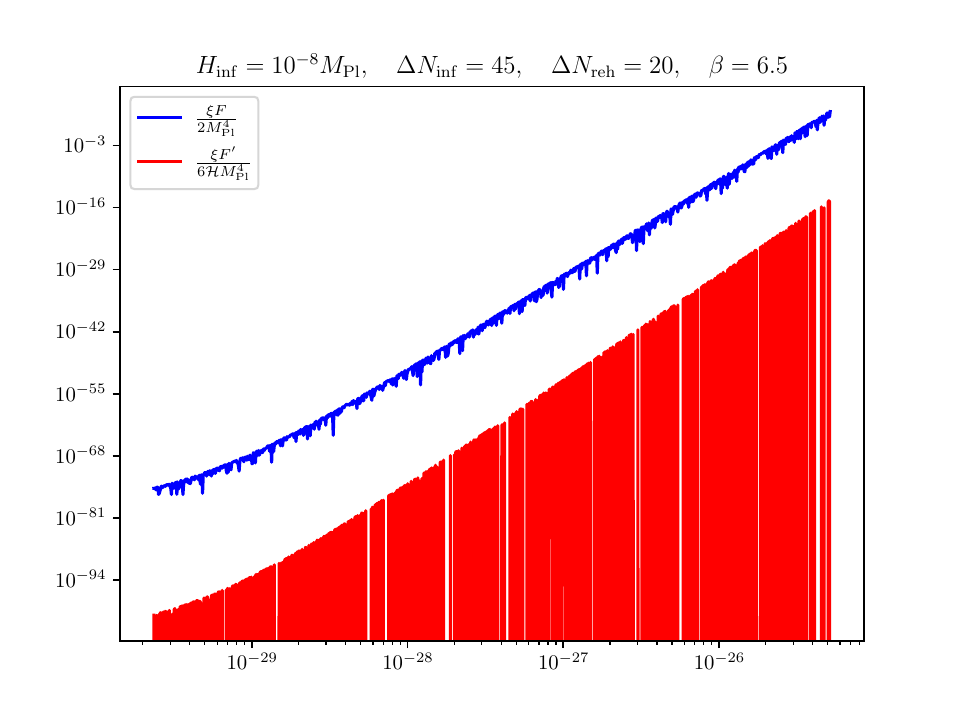}}\hfill
%\subfloat[]
{\label{b}\includegraphics[width=.5\linewidth]{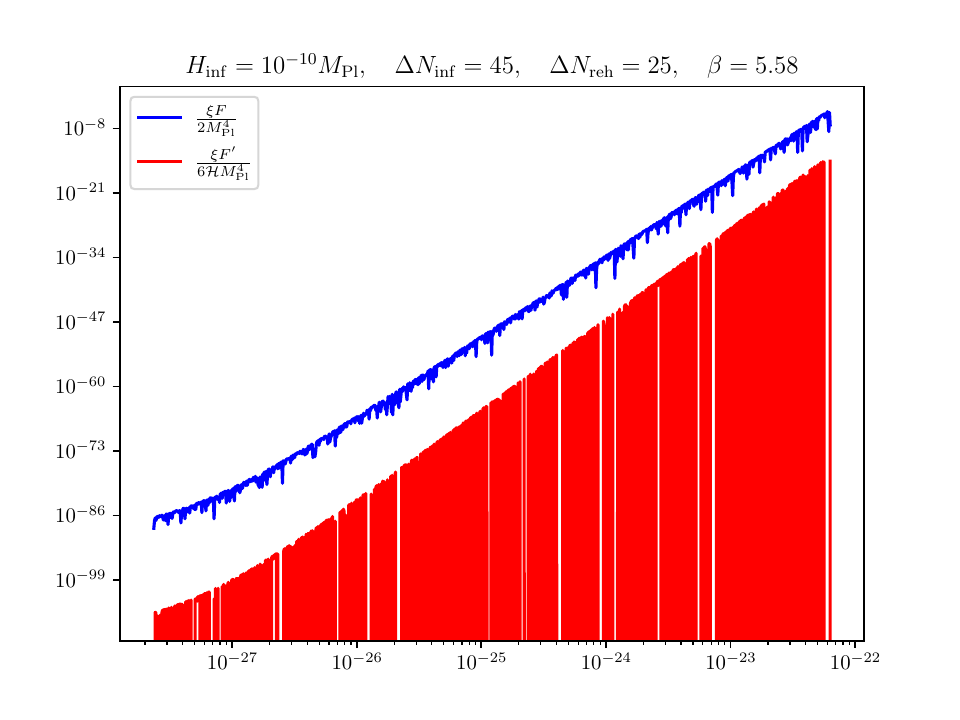}}\par 
%\subfloat[]
{\label{c}\includegraphics[width=.5\linewidth]{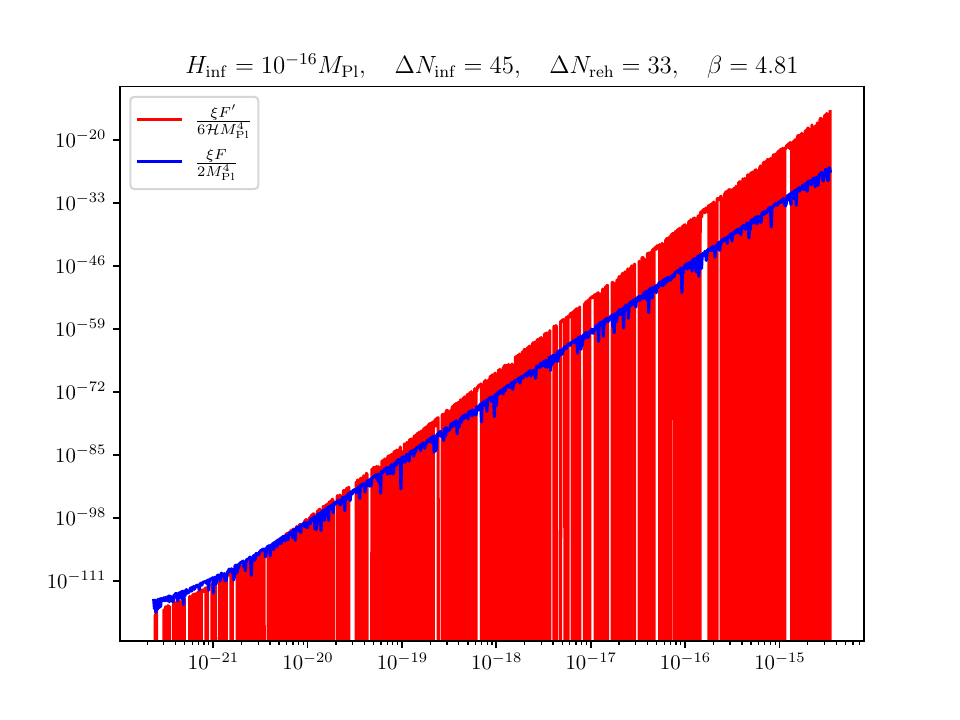}}
\caption{{\it{The terms $\frac{\xi F}{2\Mp^4}$ and $\frac{\xi F^\prime}{6\mathcal{H}\Mp^4}$ for $k=k_\mathrm{reh}$ and $\Delta N_\mathrm{inf} = 45$ by varying the parameters $H_\mathrm{inf}$, $\Delta N_\mathrm{reh}$ and $\beta$.}}}
\label{fig:xiF_terms}
\end{figure}

In \Fig{fig:xiF_terms}, we depict the $\frac{\xi F}{2\Mp^4}$ and $\frac{\xi F^\prime}{6\mathcal{H}\Mp^4}$ terms by fixing the duration of inflation at $\Delta N_\mathrm{inf}=45$ and the comoving wavenumber to $k_\mathrm{reh}$ where we expect the peak of the electromagnetically induced GW signal~\cite{Sharma:2019jtb}. As one may infer from \Fig{fig:xiF_terms}, for this particular choice of parameters both $\frac{\xi F}{2\Mp^4}$ and $\frac{\xi F^\prime}{6\mathcal{H}\Mp^4}$ terms are smaller than one and therefore can be safely neglected in the tensor modes equation of motion. For other choices of the theoretical parameters at hand $\Delta N_\mathrm{inf}$, $\Delta N_\mathrm{reh}$, $H_\mathrm{inf}$ and $\beta$, one should solve the integral-differential equation \eqref{eq:h_eq_of_motion_k_space} using advanced theoretical techniques borrowed by quantum field theory and/or numerical simulations.

\newpage
\bibliographystyle{JHEP} 
\bibliography{ref}

\end{document}